\documentclass[prl,twocolumn,superscriptaddress,showpacs]{revtex4}
\usepackage{graphicx}
\usepackage{epsfig}
\usepackage{amssymb,amsmath,amsfonts,hyperref}
\usepackage{wasysym}
\usepackage{latexsym}

\newcommand{\be}{\begin{eqnarray}}
\newcommand{\ee}{\end{eqnarray}}

\begin{document}
\title{Generalization of the singlet sector valence bond loop algorithm
to antiferromagnetic ground states with total spin $S_{tot}=1/2$}
\author{Argha Banerjee}
\affiliation{Tata Institute of Fundamental Research, 1, Homi Bhabha Rd, Colaba
Mumbai 400 005}
\author{Kedar Damle}
\affiliation{Tata Institute of Fundamental Research, 1, Homi Bhabha Rd, Colaba
Mumbai 400 005}

\begin{abstract}
We develop a generalization of the singlet sector valence bond basis projection
algorithm of Sandvik, Beach, and Evertz (A.~W.~Sandvik, Phys. Rev. Lett. 95, 207203 (2005); K.~S.~D.~Beach and A.W. Sandvik, Nucl. Phys. B750, 142 (2006); A.~W.~Sandvik and H.~G.~Evertz, arXiv:0807.0682, unpublished.)
to cases in which the ground state of an antiferromagnetic Hamiltonian
has total spin $S_{tot} = 1/2$ in a finite size system. We explain
how various ground state expectation values may be calculated by generalizations
of the estimators developed in the singlet case, and illustrate
the power of the method by calculating the ground state spin texture
and bond energies in a $L \times L$ Heisenberg antiferromagnet with
$L$ odd and free boundaries.
\end{abstract}

\pacs{75.10.Jm 05.30.Jp 71.27.+a}
\vskip2pc

\maketitle

\section{Introduction}
Understanding the ground states of strongly correlated
condensed matter systems is a central problem in computational
physics. Several approaches have had a degree of success in this
endevour. These include various sophisticated quantum monte-carlo
techniques for sampling the partition function $Tr \left(\exp(-H/T)\right) $ of a system with Hamiltonian $H$ at
temperature $T$, and using this sampling procedure to estimate
the thermal expectation values of various operators $\langle \hat{O} \rangle_T
= Tr \left( \hat{O} \exp(-{H}/T)\right)/Tr\left(\exp(-H/T)\right)$~\cite{qmc_spin_rmp,Sandvik_prb99,Prokofev_worm}.
Although these can be used for relatively large finite-size systems, they
are intrinsically finite-temperature methods and accessing the very low temperature regime involves doing calculations at successively lower temperatures
and then extrapolating.

Other approaches include various exact diagonalization techniques that obtain
the lowest energy state in a given sector. These are severally constrained
by memory requirements in terms of the system sizes they can handle. While this problem can be overcome in one dimension by the sophisticated density
matrix renormalization group method~\cite{White}, there is as yet no generalization of this method that works equally well in higher dimension, although there
has been considerable progress recently~\cite{Vidal}.

Recently, Sandvik and collaborators have developed an extremely elegant
and sophisticated projection algorithm~\cite{Sandvik_prl05,Sandvik_Beach,Sandvik_Evertz} that essentially solves the
problem of calculating the ground state expectation values of quantities
in a large class of $S=1/2$ antiferromagnetic spin systems 
which have ground state in the total spin $S_{tot} = 0$ singlet
sector. This singlet sector algorithm works in the
`bipartite valence-bond' basis for singlet states (see below), and exploits the over-completeness of this basis to develop a procedure~\cite{Sandvik_Beach} for evaluating the expectation values of some observables
${\mathcal O}$ in the (unnormalized)
ground state $|\psi_g \rangle = (-H)^m | s \rangle$ obtained by acting on an arbitrary
singlet state $| s \rangle$ with a large power of $-H$. The key to its success
is an extremely efficient~\cite{Sandvik_Evertz} procedure for stochastically sampling 
$\langle s | (-H)^{m}{\mathcal O} (-H)^m | s \rangle/\langle s | (-H)^{2m}|s \rangle$, which
allows one to handle large systems with as many as $10^4$ spins in favorable cases.

The ground state spin of a finite system made up of spin-half variables interacting
antiferromagnetically naturally depends on the nature of
the finite sample: if the total number of spin-half variables is even, one expects a ground state
in the singlet sector, while systems with an odd number of spins will have a ground state
in the total spin $S_{tot}= 1/2$ sector. For instance, an $L\times L$ square lattice
Heisenberg antiferromagnet with periodic boundary conditions and $L$ even will have a singlet
ground state, while the same magnet with $L$ odd and free boundaries will have a ground state
spin of $S_{tot}=1/2$. In many situations, it is useful to be able to handle both
kinds of finite systems. For instance, if one wants to model experiments that dope insulating
antiferromagnets with non-magnetic ions like Zn~\cite{spin_sub,defect_rmp} that substitute for the magnetic moments,
it is convenient to study $L \times L$ periodic systems with $L$ even as before, but with
one spin removed from the system to model the missing-spin defect introduced by Zn doping. 

The original valence-bond projector loop algorithm~\cite{Sandvik_prl05,Sandvik_Evertz} allows one to study the
singlet sector ground states of systems with an even number of spins. Here we ask if it is possible
to come up with an analogous procedure in the total spin $S_{tot}=1/2$ sector of systems with an odd number of spins in order
to compute properties of the $S_{tot} = 1/2$ doublet ground state of antiferromagnetic systems
with an odd number of spin-half variables interacting antiferromagnetically. As our results
demonstrate, the answer turns out to be very satisfying: Using a judiciously chosen basis for the $S_{tot}=1/2$ sector of such systems, we find that is indeed possible to construct an analogous procedure that works as well
in the total spin $S_{tot} = 1/2$ sector as the original singlet sector algorithm of Sandvik and collaborators~\cite{Sandvik_prl05,Sandvik_Beach,Sandvik_Evertz}.
Here we detail several aspects of this generalization. To illustrate the power of the method, we
also show results for the ground state `spin texture'
in the $S^z_{tot}=S_{tot}=1/2$ ground state of an $L \times L$ square lattice $S=1/2$ Heisenberg
antiferromagnet with open boundary conditions and $L$ odd and as large as $L=97$.

\section{Basis}
A judicious choice of basis is the key to generalizing the original singlet sector valence-bond projector loop
QMC algorithm to the study of bipartite spin-half antiferromagnets with $N_B$ B-sublattice sites,
$N_A=N_B+1$ A-sublattice sites, and a  doublet ground state in the $S_{tot}=1/2$ sector. While other choices may also be possible,
we find it convenient to use the basis 
\begin{eqnarray}
\left\{|{\mathcal A},a_f \sigma\rangle\right\} &\equiv&\left\{ |[{\mathcal A}(b_1) b_1],[{\mathcal A}(b_2) b_2]\dots [{\mathcal A}(b_{N_B})];a_{f}\sigma \rangle \right\} \nonumber \\
&&
\end{eqnarray}
Each member of this basis has one $A$-sublattice spin ${\mathbf S}_{a_f}$ in
either the $|a_f \sigma=\uparrow \rangle \equiv |S^z_{a_f} = +1/2\rangle$ or the $|a_f \sigma=\downarrow \rangle \equiv |S^z_{a_f} = -1/2 \rangle$ state along the quantization axis $\hat{z}$, while the $N_B$ spins ${\mathbf S}_{b_i}$ on
the $B$-sublattice sites each form a singlet state (`valence-bond') 
\begin{eqnarray}
|[{\mathcal A}(b_i) b_i]\rangle &\equiv&
\frac{\left(|{\mathcal A}(b_i)\uparrow,b_i \downarrow\rangle -|{\mathcal A}(b_i)\downarrow,b_i \uparrow\rangle \right)}{\sqrt{2}}
\end{eqnarray}
with a partner ${\mathbf S}_{{\mathcal A}(b_i)}$ on the $A$-sublattice. All basis states
are obtained by allowing all possible $a_f$, two choices for $\sigma$,
and all possible
`matching' functions ${\mathcal A}$ consistent with a given
choice of `free spin' $a_f$. Note that this basis set is actually a union of
two distinct basis sets 
\begin{eqnarray}
\left\{|{\mathcal A},a_f \uparrow \rangle\right\}&\equiv&\left\{ |[{\mathcal A}(b_1) b_1],[{\mathcal A}(b_2) b_2]\dots [{\mathcal A}(b_{N_B})];a_{f}\uparrow \rangle \right\} \nonumber \\
&&
\end{eqnarray}
and
\begin{eqnarray}
\left\{|{\mathcal A},a_f \downarrow \rangle\right\}&\equiv&\left\{ |[{\mathcal A}(b_1) b_1],[{\mathcal A}(b_2) b_2]\dots [{\mathcal A}(b_{N_B})];a_{f} \downarrow \rangle \right\} \nonumber \\
&&
\end{eqnarray}
 corresponding to the two allowed choices for the conserved quantum number $S^z_{tot}$ in
the $S_{tot}=1/2$ sector of an $SU(2)$ invariant composed of an odd number of spin-1/2s.

This basis is (over-)complete in a manner entirely analogous to
the bipartite valence-bond basis  that
was used in the original singlet sector algorithm~\cite{Sandvik_Beach,Sandvik_Evertz}. This may be seen as follows: Consider adding one extra $B$-sublattice
site $b_{N_B+1}$ to our system to make the total number of spins even. The singlet sector
of this larger system is spanned by the (over-)complete bipartite valence bond basis.
States in this basis are in one-to-one correspondence
with possible pair-wise matchings ${\mathcal P}$ that `find'
a $A$-sublattice `partner' ${\mathcal P}(b_i)$ for each $B$-sublattice
site $b_i$ to form a singlet:
\begin{eqnarray}
|{\mathcal P}\rangle & \equiv & |[{\mathcal P}(b_1) b_1],[{\mathcal P}(b_2) b_2]\dots [{\mathcal P}(b_{N_B+1}b_{N_b+1})]\rangle
\end{eqnarray}

Now, by the laws of angular momentum addition, singlet states of
the larger system can only arise from tensor products of the additional spin-half variable at site
$b_{N_b+1}$ with the $S_{tot}=1/2$ states of the smaller system. 
Therefore, to check for (over-)completeness
of our proposed basis for the smaller system,
we only need to check whether all states in the bipartite valence
bond basis of the larger system are obtainable
as tensor products of states of the additional spin ${\mathbf S}_{b_{N_b+1}}$ with
states in our proposed $S_{tot}=1/2$ basis.
This is certainly the case, as is readily seen by identifying $a_f$ with ${\mathcal P}(b_{N_b+1})$ and ${\mathcal A}$ with the restriction of ${\mathcal P}$ to
the domain $(b_1,b_2,\dots b_{N_b})$. Our proposed basis is thus overcomplete
in a manner entirely analogous to the original bipartite valence bond
basis for the singlet sector.

In practice, for $SU(2)$ symmetric spin Hamiltonians of interest to us
here, we will additionally exploit the conservation of the $z$ component of spin and
restrict attention to the basis $\left\{ |[{\mathcal A}(b_1) b_1],[{\mathcal A}(b_2) b_2]\dots [{\mathcal A}(b_{N_B})];a_{f}\uparrow \rangle \right\}$ that only spans the $S^z_{tot}=S_{tot}=1/2$ sector a system with
$N_A=N_B+1$ spin-1/2s on the $A$-sublattice, and $N_B$ spin-1/2s on the $B$
sublattice.\begin{figure}[!]
{\includegraphics[width=\columnwidth]{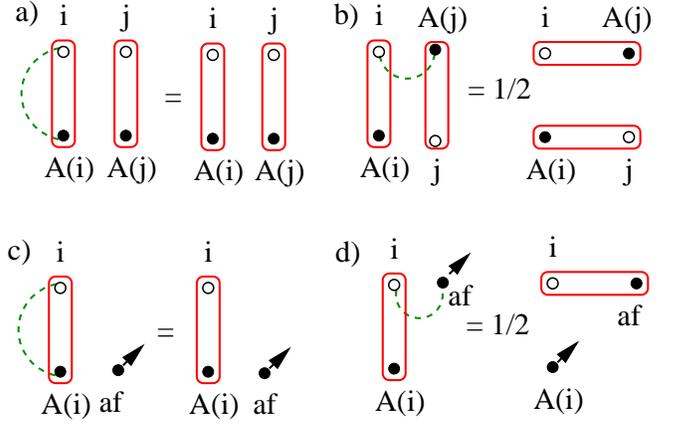}}
\caption{Action of singlet Projector (P, denoted by dashed line) in valence bond basis.}
\label{rules1}
\end{figure}

\section{Overlaps and operators}
We now indicate the changes that arise in the formulae for the wavefunction overlaps
between basis states, and for the action of exchange operators when working in the $S_{tot}=1/2$ sector.
As is well-known, the wavefunction overlap between two bipartite VB basis states  $|{\mathcal P} \rangle$ and $|{\mathcal P}'\rangle$
of the singlet-sector basis are determined by the number of cycles needed to go from the permutation ${\mathcal P}$ to the permutation
${\mathcal P}'$. More pictorially, one may consider the overlap diagram
of the two valence-bond covers viewed as `complete dimer covers' or `perfect' matchings.
This overlap diagram contains $N_{\mathrm{loop}}$ (closed) loops of various lengths $l_{\mu}$, such that each
site is part of exactly one loop (see Fig~\ref{rules1}). Knowing this overlap diagram, one may calculate the corresponding
wavefunction overlap to be $\langle {\mathcal P}|{\mathcal P}' \rangle
= 2^{N_{\mathrm{loop}}}/2^{N_{s}/2}$, where $N_{s}$, the total number of spins is
assumed even.

Generalizing this to states in our basis for the $S_{tot}=1/2$ sector, we note that the corresponding picture is now terms
of the overlap diagram of two {\em partial valence bond covers} $({\mathcal A},a_f)$ and $({\mathcal A}',a'_f)$, each
of which leaves one site free (uncovered by a valence bond). Such an overlap diagram necessarily involves
exactly  one `open string' of length $l_f$ connecting $a_f$ to $a'_f$, in addition to $N_{\mathrm{loop}}$ (closed) loops of various lengths $l_{\alpha}$ (see Fig~\ref{rules1}). An elementary calculation reveals that the
wavefunction overlap  in our $S_{tot}=1/2$ case is given as $\langle {\mathcal A}a_f\sigma|{\mathcal A}' a'_f \sigma'\rangle
= \delta_{\sigma \sigma'}2^{N_{\mathrm{loop}}}/2^{(N_{s}-1)/2}$, with $N_{s}$, the number of sites,  now taken to be odd.

The original singlet sector algorithm relies heavily~\cite{Sandvik_prl05,Sandvik_Beach,Sandvik_Evertz} on a particularly simple action of
operators $P_{ij} = \eta_i \eta_j{\mathbf S}_i \cdot {\mathbf S}_j +1/4$ on basis states $|{\mathcal P}\rangle$.
Here $\eta_i = +1$ ($\eta_i = -1$) for $i$ belonging to the $A$-sublattice ($B$-sublattice), and thus,
the operator $P_{a_\alpha b_\beta}$ that connects an $A$-sublattice site $a_\alpha$ to a $B$-sublattice site $b_\beta$ is precisely the projection operator that projects to the singlet state of the two spins ${\mathbf S}_{a_\alpha}$
and ${\mathbf S}_{b_\beta}$. Our key observation, which allows us to generalize this algorithm to the
the $S_{tot}=1/2$ case, is that the action of $P_{ij}$ on states in our $S_{tot}=1/2$ basis remains simple.
This is seen as follows: If neither $i$ nor $j$ correspond to the `free' spin, $P_{ij}$ acts
exactly as in the earlier singlet sector case (Fig~\ref{rules1}(a,b)):
\begin{widetext}
\begin{eqnarray}
\nonumber
 {P_{{\mathcal A}(b_\alpha) b_\alpha}} |...[{\mathcal A}(b_\alpha) b_\alpha]...\; ; a_f \uparrow\rangle &=& 
|...[{\mathcal A}(b_\alpha),b_\alpha]... \; ; a_f \uparrow\rangle,\\ \nonumber
 {P_{{\mathcal A}(b_\alpha) b_\beta}} |...[{\mathcal A}(b_\alpha),b_\alpha]...[{\mathcal A}(b_\beta),b_\beta]...\; ;a_f \uparrow\rangle &=& 
\frac{1}{2} |...[{\mathcal A}(b_\alpha),b_\beta]...[{\mathcal A}(b_\beta),b_\alpha]...\; ;a_f \uparrow\rangle,\\ \nonumber
 {P_{{\mathcal A}(b_\alpha) {\mathcal A}(b_\beta)}}
|...[{\mathcal A}(b_\alpha),b_\alpha]...[{\mathcal A}(b_\beta),b_\beta]...\; ;a_f \uparrow\rangle &=& \frac{1}{2}
|...[{\mathcal A}(b_\alpha),b_\beta]...[{\mathcal A}(b_\beta),b_\alpha]...\; ;a_f \uparrow\rangle,\\\nonumber
 {P_{b_\alpha b_\beta}} |...[{\mathcal A}(b_\alpha),b_\alpha]...[{\mathcal A}(b_\beta),b_\beta]...\; ;a_f \uparrow\rangle &=& 
\frac{1}{2} |...[{\mathcal A}(b_\alpha),b_\beta]...[{\mathcal A}(b_\beta),b_\alpha]...\; ;a_f \uparrow\rangle.\\
\end{eqnarray}
\end{widetext}
On the other hand, if either $i$ or $j$ correspond to the free site $a_f$, one can easily check that the following holds (Fig~\ref{rules1}(c,d)):
\begin{widetext}
\begin{eqnarray}
\nonumber 
 {P_{a_f b_\alpha}}|...[{\mathcal A}(b_\alpha) b_\alpha]...\; ; a_f \uparrow\rangle &=& \frac{1}{2}|...[a_f,b_\alpha]...\; ; {\mathcal A}(b_\alpha)\uparrow\rangle,\\ \nonumber
 {P_{a_f {\mathcal A}(b_\alpha})}|...[{\mathcal A}(b_\alpha) b_\alpha]...\; ; a_f \uparrow\rangle &=& \frac{1}{2}|...[a_f,b_\alpha]... \; ; {\mathcal A}(b_\alpha)\uparrow\rangle.\\
\label{reconnectionamplitude}
\end{eqnarray}
\end{widetext}
Thus $P_{ij}$ either causes no
change or rearranges exactly one pair of valence bonds to give a
new basis state with amplitude $1/2$, or reconnects one valence
bond to move the free spin to give a new basis state, again with amplitude $1/2$. The important thing to note is that these rules are in complete analogy
to the original singlet sector case.
 
\begin{figure}[!]
{\includegraphics[width=\columnwidth]{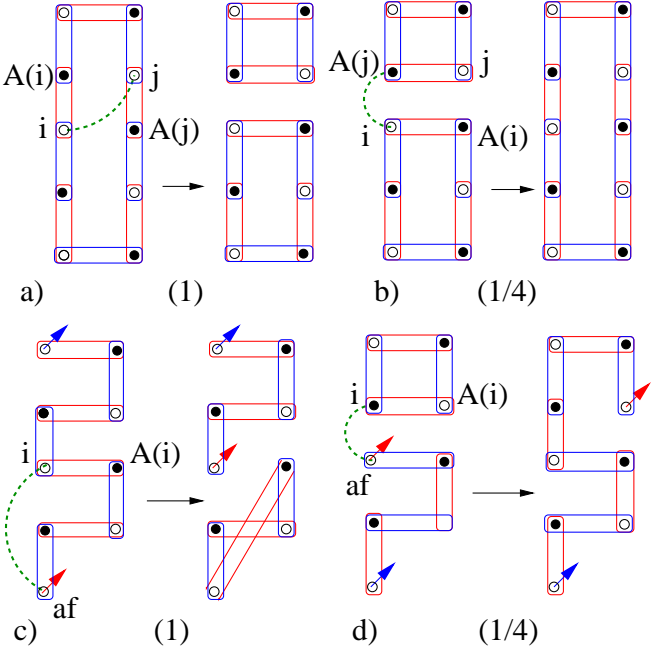}}
\caption{Action of a singlet projector on the overlap loops for closed loops (a, b) and open string (c,d). Corresponding weights are given below within parenthesis.}
\label{Pijrules}
\end{figure}

By analogy to the original singlet sector work~\cite{Sandvik_prl05,Sandvik_Beach,Sandvik_Evertz}, this allows us
to formulate a convenient prescription for the calculation
of $\langle {\mathcal A}' a'_f \uparrow |P_{ij}| {\mathcal A} a_f \uparrow\rangle$ between two of our basis states by writing  $\langle {\mathcal A}' a'_f \uparrow |P_{ij}| {\mathcal A} a_f \uparrow\rangle = W_{ij}\langle {\mathcal A}' a'_f \uparrow | {\mathcal A} a_f \uparrow\rangle$ and developing rules
for the weight $W_{ij}$ by comparing the overlap diagram
of $({\mathcal A}' a'_f \uparrow )$ and $(P_{ij}| {\mathcal A} a_f \uparrow)$ with the original overlap diagram of $({\mathcal A}' a'_f \uparrow )$ and $({\mathcal A} a_f \uparrow)$: If the action of $P_{ij}$ makes no changes in the
original overlap diagram,
$W_{ij} = 1$. In addition, $W_{ij} = 2 \times 1/2 = 1$ if a loop is split into two loops or the open string is split into
one loop and another open string; here, the factor of two comes from the fact that the number of loops in
the overlap diagram increases by one, while the factor of half has its origins in the reconnection amplitude of
one-half in Eqn~\ref{reconnectionamplitude}. On the other hand, if  two loops fuse into one, or if the open
string fuses with a loop to give a larger open string, then $W_{ij}= (1/2) \times (1/2)$, where the first
factor of half reflects the fact that the number of loops is reduced by one, while the second factor
of one-half comes from the reconnection amplitude in Eqn~\ref{reconnectionamplitude}. These rules
are tabulated in Fig~\ref{Pijrules}, and the important thing to note is that the open string can
be treated on equal footing with (closed) loops in all cases, allowing one to generalize the singlet
sector rules directly to the $S_{tot} =1/2$ sector case discussed here.  

\section{Generalization of the Sandvik-Evertz algorithm}

With all of this in hand, it is now easy to see that the singlet sector
algorithm of Sandvik and collaborators~\cite{Sandvik_prl05,Sandvik_Beach,Sandvik_Evertz} generalizes straightforwardly to the $S_{tot}=1/2$ sector case for
Hamiltonians of the form $H = -\sum_b H_b$, where each piece $-H_b$ of the Hamiltonian is
a projector $P_{b_1b_2}$ acting on bond  $b$ connecting spins $b_1$ and $b_2$.
We start with an arbitrary $S_{tot}=1/2$ $S^z_{tot}=1/2$ state $|\psi_{1/2 \uparrow}\rangle$, say $|\psi_{1/2\uparrow}\rangle = | {\mathcal A} a_f \uparrow\rangle$. As in the original singlet sector algorithm, we wish to stochastically sample $\langle \psi_{1/2 \uparrow}|(-H)^{2m}|\psi_{1/2\uparrow}\rangle$ by sampling all possible operator strings $\prod_{\tau=1}^{2m}H_{b_\tau}$
in the decomposition of $(-H)^{2m}$ into a sum of products, 
with weight for each such string being proportional to  $\langle \psi_{1/2 \uparrow}|\prod_{\tau=1}^{2m}H_{b_\tau}|\psi_{1/2 \uparrow}\rangle$. To do
this, one splits each $H_{b}$ into a term $H_{b}^{\sigma=d}$ that is diagonal in the $\{S^z_i\}$ eigenbasis, and a term $H_{b}^{\sigma=o}$
that is offdiagonal. In addition, one writes $|\psi_{1/2 \uparrow}\rangle$ in this basis as $|\psi_{1/2 \uparrow}\rangle = \sum_{\{S^z_i\}}C_{\{S^z_i\}\uparrow}|\{S^z_i\}\rangle$.
As in the original singlet
sector case, each term $C_{\{{S^z_i}'\}\uparrow}C_{\{S^z_i\}\uparrow}\langle \{{S^z_i}'\}|\prod_{\tau=1}^{2m}H^{\sigma_\tau}_{b_\tau}|\{S^z_i\}\rangle$ is generated by working in a `spacetime' loop representation and using a combination
of `diagonal updates' in which some $H^{\sigma_\tau = d}_{b_\tau}$ is moved to a different bond $b_{tau}'$
and `loop updates' whereby each space-time loop is flipped with probability half; this loop update allows one to switch between diagonal and off-diagonal
pieces of a given set of bond operators, while simultaneously sampling
all possible spin configurations $\left\{ S^z_i\right\}$ in the state
at $\tau=0$ and $\tau=2m$. The only
difference with the original singlet sector case is that we now have {\em precisely one open string in the space-time
diagram}, which connects the `free spin at $\tau = 0$', {\em i.e} the unpaired spin in $|\psi_{1/2}\uparrow\rangle$ to
the `free spin at $\tau=2m$', {\em i.e} the unpaired spin in $\langle \psi_{1/2}\uparrow|$, and which
{\em cannot be flipped}, since the unpaired spins in all our $S_{tot}=1/2$, $S^z_{tot}=1/2$ basis states have fixed $z$ projection of $+1/2$.
Finally, as in the singlet case, we can easily generalize this procedure
to treat Hamiltonians that also contain products $-P_{b} P_{b'}$
of projectors $P$ acting on distinct bonds $b$ and $b'$ of the lattice.

\begin{figure}[!]
{\includegraphics[width=\columnwidth]{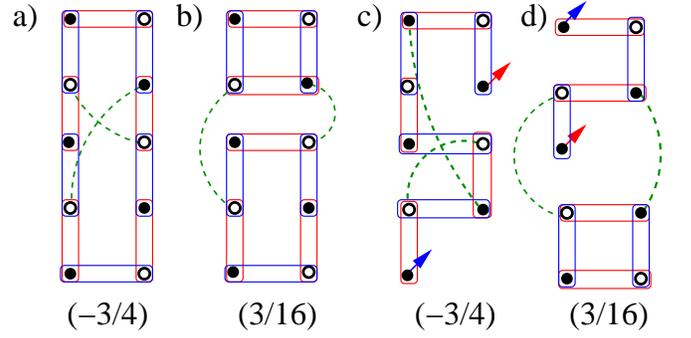}}
\caption{Overlap graphs contributing to the estimator of $\langle({\vec{m}^2})^2\rangle \equiv \sum_{ijkl} (P_{ij} -1/4)(P_{kl}-1/4) \rangle $ with corresponding values of $W_{ijkl} - W_{ij}W_{kl}$ (defined in text) given within parenthesis.}
\label{ssssrules}
\end{figure}
\section{Estimators}
As in the singlet sector case~\cite{Sandvik_prl05,Sandvik_Beach,Sandvik_Evertz}, physical properties can be calculated by taking each space-time
loop diagram generated by the algorithm and `cutting it at $\tau=m$' to obtain the overlap
diagram that represents the overlap of $\prod_{\tau=1}^{m}H^{\sigma_\tau}_{b_\tau}|\psi_{1/2 \uparrow}\rangle$
with $\langle \psi_{1/2 \uparrow}\prod_{\tau=m+1}^{2m}H^{\sigma_\tau}_{b_\tau}$.

Consider for instance the Neel order parameter $\vec{m} = \sum_i \eta_i\vec{S}_i$. Clearly, $\langle m_x \rangle = \langle m_y \rangle = 0$. However,
$\langle m_z \rangle $ receives contributions from sites on the open string
in the overlap diagram, since the open string, in contrast to (closed) loops,
has only one orientation and therefore cannot be flipped. More formally,
we may write $\langle {\mathcal A}' a'_f \uparrow |S^z_i| {\mathcal A} a_f \uparrow\rangle = W_i\langle {\mathcal A}' a'_f \uparrow | {\mathcal A} a_f \uparrow\rangle$ and note that $W_i = \eta_i/2$ if $i$ is part of the open string
in the overlap diagram between $({\mathcal A}' a'_f)$ and $({\mathcal A} a_f)$ 
and $0$ otherwise. We thus find
$\langle m_z \rangle  = \langle l_f \rangle$, where the angular brackets on the right denote the ensemble average
over the ensemble of overlap diagrams generated by the modified
$S_{tot} =1/2 S_{tot}^z = 1/2$ sector algorithm outlined above.

We now turn to 
$\langle \vec{m}^2\rangle = \langle\sum_{ij}(P_{ij} -1/4)\rangle$.
As noted earlier, whenever $i$ and $j$ are both in the open string
or the same (closed) loop,
the corresponding weight $W_{ij} = 1$, while $W_{ij} =1/4$ when $i$ and $j$
do not both belong to the open string or the same (closed) loop. For an
overlap diagram with closed loops of lengths $l_\alpha$ (with $\alpha = 1,2 \dots N_l-1$) and an open
string of length $l_{N_{l}}\equiv l_f$, the latter can occur in $\sideset{}{'}\sum_{\alpha, \beta = 1}^{ N_l}l_\alpha l_\beta$ ways where the prime on the sum indicates that
$\alpha = \beta$ is disallowed, while the former can occur in $\sum_{\alpha=1}^{N_l} l_\alpha^2$
ways. As in the singlet sector case, we thus obtain
$\langle \vec{m}^2 \rangle = \langle \frac{1}{4}\sideset{}{'}\sum_{\alpha, \beta = 1}^{ N_l}l_\alpha l_\beta +\sum_{\alpha=1}^{N_l} l_\alpha^2 - \frac{1}{4}\sum_{\alpha,\beta=1}^{N_l}l_\alpha l_\beta$, where the angular brackets on the right indicate average over the
ensemble of overlap diagrams generated by the algorithm.
This reduces to
\begin{eqnarray}
\langle m^2 \rangle & = &\langle \frac{3}{4}\sum_{\alpha=1}^{N_l} l_\alpha^2\rangle
\end{eqnarray}
where the angular brackets on the right again denote averaging over
the ensemble of overlap diagrams generated by the algorithm, and the important thing to note is that this estimator treats
the open string ($\alpha=N_l$) on the same footing as the closed loops
($\alpha=1,2,\dots N_l-1$).

Finally, we consider the ground state expectation value of the fourth power of the Neel order parameter, {\em i.e}
$\langle (\vec{m}^2)^2\rangle$. To derive the estimator for this in the $S_{tot}=1/2$, $S^z_{tot} =1/2$, we
follow Sandvik and Beach~\cite{Sandvik_Beach}, and write $\langle (\vec{m}^2)^2\rangle = \langle \sum_{ij} \sum_{kl} (P_{ij} -1/4)(P_{kl} - 1/4)\rangle$. As in Ref~\onlinecite{Sandvik_Beach}, we note that the estimator for this quantity differs from
the square of the estimator for $\vec{m}^2$ only when the action of $P_{ij}$ `interferes' with the action
of $P_{kl}$, {\em i.e} when the actual weight $W_{ijkl} \equiv \langle {\mathcal A}' a'_f \uparrow |P_{ij}P_{kl}| {\mathcal A} a_f \uparrow\rangle/\langle {\mathcal A}' a'_f \uparrow | {\mathcal A} a_f \uparrow\rangle$ differs
from the product $W_{ij} W_{kl}$ of the independent weights $W_{ij} \equiv \langle {\mathcal A}' a'_f \uparrow |P_{ij}| {\mathcal A} a_f \uparrow\rangle/\langle/{\mathcal A}' a'_f \uparrow | {\mathcal A} a_f \uparrow\rangle$ and $W_{kl}$ (defined
analogously to $W_{ij}$). As in the singlet sector case, this happens only in the two cases shown in Fig~\ref{ssssrules},
where the difference $W_{ijkl} - W_{ij}W_{kl}$ has been tabulated. Thus, the only new
calculation needed is a count of the number of ways in which each of the cases Fig~\ref{ssssrules} (a), (b), (c), (d)
arise, weighted by the corresponding values of $W_{ijkl} - W_{ij}W_{kl}$. It is at this step that
the open string needs to be treated separately, since we find that this count for a open string in Fig~\ref{ssssrules} (c) differs
from the analogous count for a closed loop in Fig~\ref{ssssrules} (a) by precisely one: Fig~\ref{ssssrules} (a)
can arise in $\frac{1}{3}l_\alpha^4 - \frac{4}{3}l_\alpha^2$ ways, while Fig~\ref{ssssrules} (c) can arise
in $\frac{1}{3}l_f^4 - \frac{4}{3}l_f^2+1$ ways. On the other hand, both Fig~\ref{ssssrules} (b) and (d) arise
in precisely $2 l_\alpha^2 l_\beta^2$ ways (with $l_\beta \equiv l_f$) for Fig~\ref{ssssrules} (d).

With all this in hand, we obtain
\begin{eqnarray}
\langle (\vec{m}^2)^2 \rangle & = & \left(\frac{3}{4}\sum_{\alpha=1}{N_l}l_\alpha^2\right)^2  + \nonumber \\
&&\frac{6}{16}\sideset{}{'}\sum_{\alpha, \beta = 1}^{ N_l} l_\alpha^2 l_\beta^2 - \frac{1}{4}\sum_{\alpha=1}^{N_l}(l_\alpha^4-4l_\alpha^2)
-\frac{3}{4} \; ,
\end{eqnarray}
which reduces to .
\begin{eqnarray}
\nonumber
\langle (\vec{m}^2)^2 \rangle &=&\sum_{\alpha=1}^{N_l}{( -\frac{5}{8} l_\alpha^4 + l_\alpha^2 )} + \frac{15}{16}(\sum_{\alpha=1}^{N_l}{l_\alpha^2})^2   -\frac{3}{4}.\\
\end{eqnarray}

Again, the thing to note is that the presence of the open string only changes the estimator by an addition
constant $-\frac{3}{4}$ when compared to the corresponding expression in the singlet sector case~\cite{Sandvik_Beach}.


\begin{figure}[!]
{\includegraphics[width=\columnwidth]{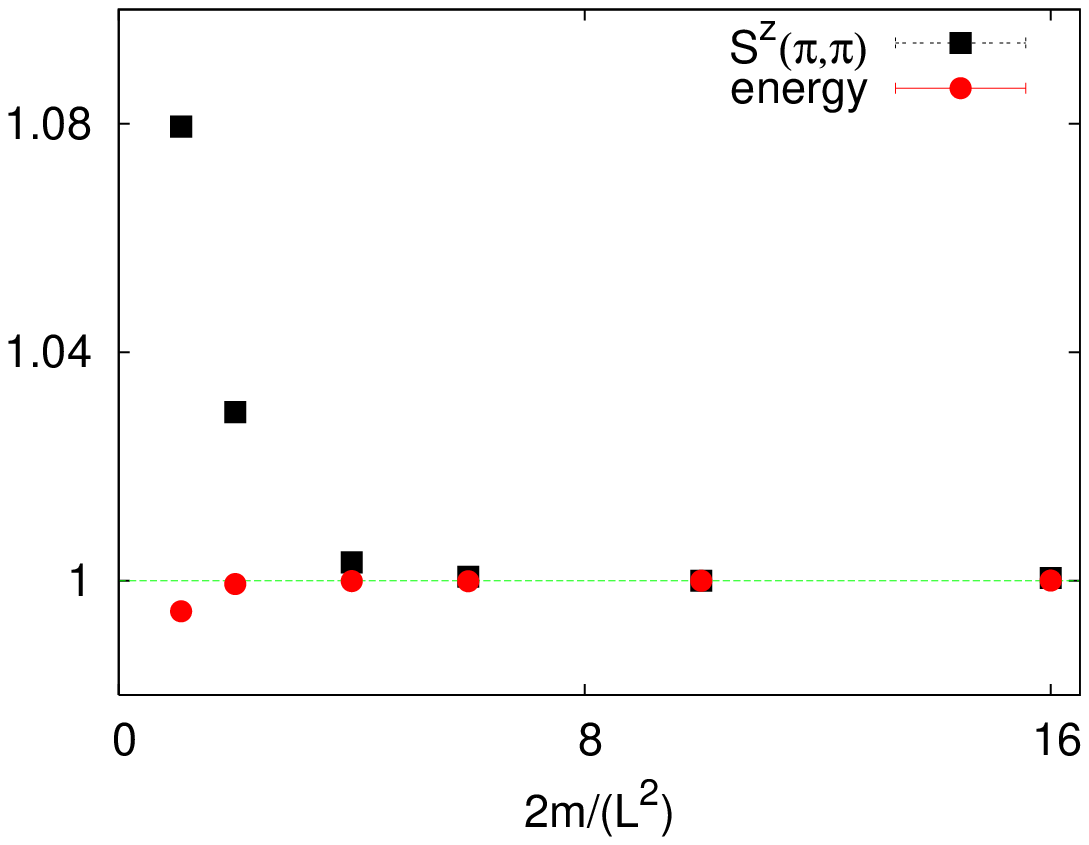}}
{\includegraphics[width=\columnwidth]{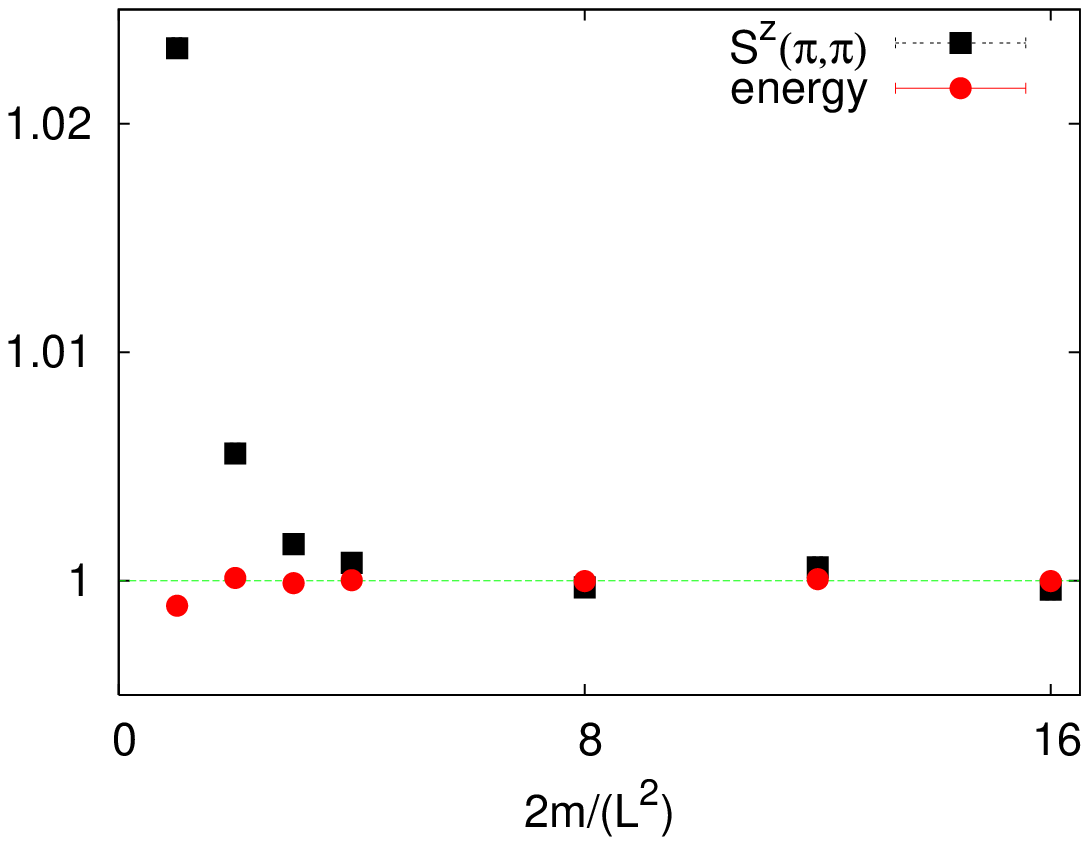}}
\caption{Variation of the ratio of Monte Carlo estimates of $S_z(\pi, \pi)$ and energy to exact values as a function of  projection length: for a $3\times5$ open system (top) and a $4\times4$ periodic system (below).}
\label{opensaturation}
\end{figure}
\begin{figure}[!]
{\includegraphics[width=\columnwidth]{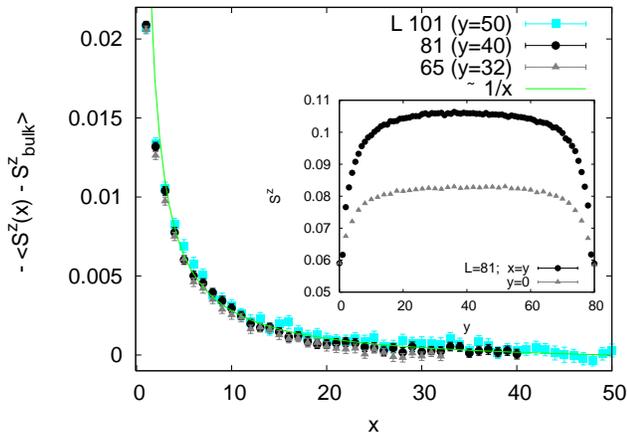}}
\caption{Boundary induced spin texture decays as $\sim 1/x$ to the bulk value along a direction perpendicular to the boundary: negative differences from bulk value shown along a cut perpendicular to the boundary for  various systemsizes. Inset shows the $<S^z(x,y)>$ texture on the A-sublattice sites along a diagonal and an edge of a $81\times81$ system.}
\label{opensz}
\end{figure}
\section{Illustrative results and outlook}
By way of illustration, we show results for a $L_x \times L_y$
square lattice Heisenberg antiferromagnet with $L_x$, $L_y$ odd and open
boundary conditions, and for a $L \times L$ square
lattice Heisenberg antiferromagnet with periodic boundary
conditions and $L$ even, but with one site missing.

To benchmark the method, we first compare the results for $L_x=3$, $L_y=5$ open
boundary condition system and a $L=4$ period boundary condition system having
one site missing with the corresponding exact diagonalization results. In Fig~\ref{opensaturation}, we
show the dependence of the estimators for ground state energy, $S_z(\pi, \pi)$ and
$\vec{m}_s^2$ as a function of the projection power $m$; this performance is comparable
to the performance of the original algorithm in the singlet sector, and thus
our modification provides a viable method to study antiferromagnets forced to have
a $S_{tot}=1/2$ ground state due to the nature of the finite sample.

With this in hand, we move on to some illustrative physics results. For purposes of illustration,
we consider a large open boundary condition system with an odd number of sites. In
Fig~\ref{opensz} shows the magnetization at all sites of A sublattice for a open system with sizes $L=$.
 As noted by Metlitski and Sachdev~\cite{Metlitski_Sachdev}, the effect of the boundary is to decrease the sublattice magnetization
near it  which is then restored to 
its bulk value away from boundary in a power-law manner~\cite{Hoglund_Sandvik,Metlitski_Sachdev}. As demonstrated
by Ref~\onlinecite{Metlitski_Sachdev},
this suppression goes away as a power law $1/|\vec{r}|$ as a function of
distance $|\vec{r}|$ from the edge. Using our method, we can directly
calculate $\langle S(\vec{r})\rangle$ in an odd by odd square lattice. On general
grounds, one
expects that $(-1)^{\vec{r}}\langle S(\vec{r})\rangle$ will also obey this prediction
of Metlitski and Sachdev, although this quantity is not directly related to the
usual definition of the Neel order parameter. With this in mind, we compare
our results with the predictions from Ref~\onlinecite{Metlitski_Sachdev}, 
and find extremely good agreement, pointing to the usefulness of our approach.

\section{Acknowledgements}
We acknowledge computations resources of TIFR, and funding from DST-SRC/S2/RJN-25/2006.


\begin{thebibliography}{999}
\bibitem{Sandvik_prl05}A.~W.~Sandvik, Phys. Rev. Lett. 95, 207203 (2005).


\bibitem{Sandvik_Beach}K.~S.~D.~Beach and A.W. Sandvik, Nucl. Phys. B750, 142 (2006).

\bibitem{Sandvik_Evertz} A.~W.~Sandvik and H.-G.~Evertz, arXiv:0807.0682, unpublished.

\bibitem{Sandvik_prb99}A. W. Sandvik, Phys. Rev. B 59, R14157 (1999).

\bibitem{qmc_spin_rmp}E. Manousakis, Rev. Mod. Phys. 63, 1 (1991).

\bibitem{Prokofev_worm}N. V. Prokof'ev, B. V. Svistunov, and I. S. Tupitsyn, Pis'ma Zh. Eksp. Teor. Fiz. 64, 853 (1996)[Sov. Phys. JETP Lett. 64, 911 (1996)].
\bibitem{White}S.~R.~White,Phys. Rev. Lett. 69, 2863(1992).
\bibitem{Vidal}G. Evenbly and G. Vidal, Phys. Rev. Lett. 102, 180406 (2009).  
\bibitem{defect_rmp}H. Alloul, J. Bobroff, M. Gabay, and P. Hirschfeld, Rev. Mod. Phys. 81, 45 (2009).

\bibitem{spin_sub}O. P. Vajk, P. K. Mang, M. Greven, P. M. Gehring, and J. W. Lynn, 2002, Science 295, 1691.

\bibitem{Hoglund_Sandvik}K. H. Huglund, and A W Sandvik, Phys. Rev. B 79, 020405 (2009).

\bibitem{Metlitski_Sachdev}M. A. Metlitski and S. Sachdev, Phys. Rev. B 77, 054411 (2008)
\end{thebibliography}
\end{document}